\documentclass[twocolumn,twoside,slac_two]{revtex4}
\usepackage{graphicx}
\usepackage{fancyhdr}
\begin{document}

\title{Searches for high energy solar flares with \textsl{Fermi}-LAT}

\author{G. Iafrate}
\affiliation{INAF - Astronomical Observatory of Trieste, Italy and
INFN Trieste, Italy}
\author{F. Longo}
\affiliation{INFN Trieste, Italy and Dipartimento di Fisica,
Trieste, Italy}
\author{on behalf of the Fermi Large Area Telescope Collaboration}
\affiliation{}

\begin{abstract}
The \textsl{Fermi} Large Area Telescope (LAT) has been surveying the
sky in gamma rays from 30 MeV to more than 300 GeV since August
2008. \textsl{Fermi} is the only mission able to detect high energy
$(> \text{few hundreds MeV})$ emission from the Sun during the new
solar cycle 24: the Solar System Science Group of the \textsl{Fermi}
team is continuously monitoring high energy emission from the Sun
searching for flare events. Preliminary upper limits $(>100\text{
MeV})$ have been derived for all solar flares detected so far by
other missions and experiments (RHESSI, \textsl{Fermi} GBM, GOES).
Upper limit for flaring Sun emission (integrated over one year of
data) has also been derived. Here we present the analysis techniques
as well as the details of this search and the preliminary results
obtained so far.
\end{abstract}

\maketitle

\thispagestyle{fancy}

 \begin{figure*}[th]
  \centering
  \includegraphics[width=5in]{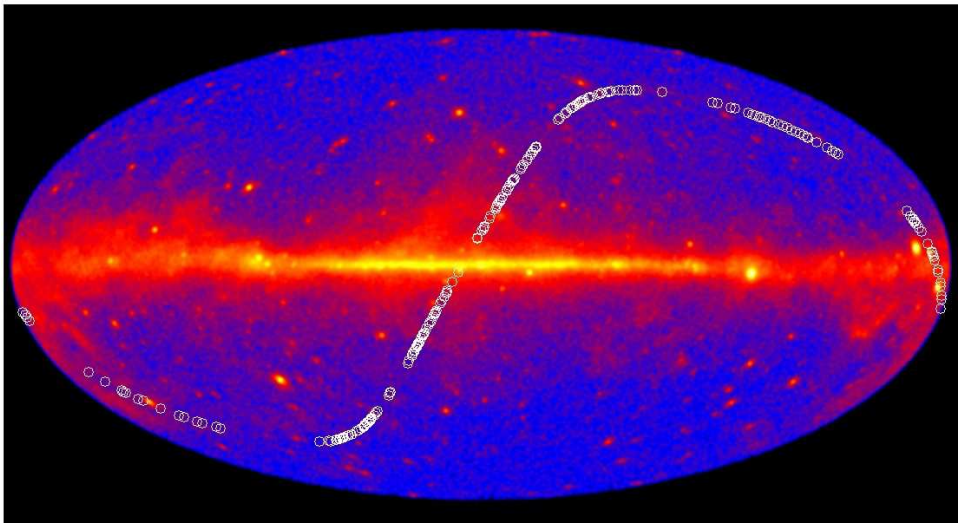}
  \caption{Solar flares detected by RHESSI and analysed in this search,
  superimposed on a count map of LAT data ($E>200\textnormal{ MeV}$) for a easier localization in the sky. There is no evidence of
  correlation between flare positions and excesses of the LAT events.}
  \label{cmap}
 \end{figure*}

\section{Introduction}
\emph{Fermi} was successfully launched from Cape Canaveral on 2008
June 11. It is currently in an almost circular orbit around the
Earth at an altitude of 565 km having an inclination of
25.6$^{\circ}$ and an orbital period of 96 minutes. After an initial
period of engineering data taking and on-orbit
calibration\cite{Fermi}, the observatory was put into a sky-survey
mode in August 2008. The observatory has two instruments onboard,
the Large Area Telescope (LAT)\cite{LAT}, a pair-conversion
gamma-ray detector and tracker (energy range 30 MeV - $>300$ GeV)
and a Gamma-ray Burst Monitor (GBM), dedicated to the detection of
gamma-ray bursts (energy range 8 keV - 40 MeV). The instruments on
\emph{Fermi} provide coverage over the energy range measurements
from few keV to several hundreds of GeV.

Here we report Fermi LAT limits on emission $> 100 \text{ MeV}$ for
the few flares detected by other missions over the past year. Solar
flares are the most energetic phenomena that occur within our Solar
System. A flare is characterized by the impulsive release of a huge
amount of energy, previously stored in the magnetic fields of active
regions. During a flare plasma of the solar corona and chromosphere
is accelerated and electromagnetic radiation covering the entire
spectrum is emitted. The production of $\gamma$-rays involves
flare-accelerated charged-particle (electrons, protons and heavier
nuclei) interactions with the ambient solar atmosphere. Electrons
accelerated by the flare, or from the decay of $\pi^{\pm}$
secondaries produced by nuclear interactions, yield X and
$\gamma$-ray  bremsstrahlung radiation with a spectrum that extends
to the energies of the primary particles. Proton and heavy ion
interactions also produce $\gamma$-rays through $\pi^{0}$ decay,
resulting in a spectrum that has a maximum at 68 MeV\cite{share}.

The frequency of solar flares follows the 11 year solar activity
cycle. Most intense flares occur during the maximum, but intense
flares can occur also in the rising and decreasing phases of the
cycle. The new solar activity cycle 24 has started at the beginning
of year 2008, the maximum is predicted in year 2012. \emph{Fermi}
has been launched during the minimum of the solar cycle, so the
frequency and the intensity of solar flares will increase throughout
most of the mission. If the goal of a 10-year mission life is
achieved, \emph{Fermi} will operate for nearly the entire duration
of solar cycle 24. During this time, \emph{Fermi} will be the only
high-energy observatory ($>\text{few hundreds MeV}$) to complement
several solar missions at lower energies: RHESSI, GOES, SoHO,
Coronas.

\section{Previous observations}

The 2005 January 20 solar flare produced one of the most intense,
fastest rising and hardest solar energetic particle events ever
observed in space or on the ground. $\gamma$-ray measurements of the
flare\cite{share06}\cite{grechnev} revealed what appear to be two
separate components of particle acceleration at the Sun: i) an
impulsive release lasting $\sim10$ min with a power-law index of
$\sim3$ observed in a compact region on the Sun and, ii) an
associated release of much higher energy particles having a spectral
index $\leq2.3$ interacting at the Sun for about two hours.
Pion-decay $\gamma$-rays appear to dominate the latter component.
Such long-duration high-energy events have been observed before,
most notably on 1991 June 11 when the EGRET instrument on CGRO
observed $>50$ MeV emission for over 8 hours\cite{kanbach}. It is
possible that these high-energy components are directly related to
the particle events observed in space and on Earth.

\emph{Fermi} will improve our understanding of the mechanisms of the
$\gamma$-ray emission by solar flares thanks to its large effective
area, sensitivity and high spatial and temporal resolution.

\section{Monitor of solar cycle 24}

The solar cycle 24 has started at the beginning of 2008, but
actually we are in an extended period of minimal solar activity. We
are seeing an interesting diminished level of activity. There are
some discussion ongoing if sunspots and flares ever return and how
unusual is this behavior\cite{nugget}. A closer look at the daily
values of three indices: F10.7 (10 cm radio flux from the Sun), the
total solar irradiance TSI, and the classical sunspot number give
only a little appearance of a up-turn. In the modern era there is no
precedent for such a protracted activity minimum, but there are
historical records from a century ago of a similar pattern
(transition between cycles 13 and 14, 107 years ago).

Activity is expected to pick up in the next months. In the meanwhile
is a good opportunity to use the excellent data available from many
satellites to improve LAT analysis of solar flares and practise in
flare monitoring and analysis, to be ready when the first intense
flare of cycle 24 will arrive.

\section{Data selection}

Since August 2008 flares detected by RHESSI and GOES have been
continuously monitored, analysing LAT data for flare events
potentially detectable by the LAT and computing upper limits on the
solar high energy emission. Solar flares have been searched in LAT
data from Augusr 2008 to the end of August 2009. LAT data have been
analysed in the time intervals of flares detected by GBM, RHESSI and
GOES. A zenith cut of $105^{\circ}$ has been applied to eliminate
photons from the Earth's albedo. For this analysis the \lq\lq
Diffuse\rq\rq\ class\cite{LAT} selection has been adopted,
corresponding to the events with the highest photon classification
probability, using the IRFs (Instrumental Response Functions)
version P6\_V3.

\section{Analysis method}

The list of flares detected by RHESSI\cite{rhessi} and the
\emph{Solar Monitor} web site\cite{solarmonitor} were monitored
constantly at a daily basis.  Flares seen by RHESSI and GOES with
more than $10^{5}$ counts (detected by RHESSI) have been selected.
For each of these flares start and end time of the event in
\emph{Fermi} MET (Mission Elapsed Time), the position of the Sun
during the flare and the angle of the Sun direction with the LAT
boresight have been computed.

The excess of events in the LAT data has been searched for flares
within the LAT field of view (angle with the LAT boresight
$<80^{\circ}$). Although the Sun is a moving source in the sky,
covering about $1^{\circ}$ per day, in this analysis the Sun has
been considered as a fixed source, due to the short duration of the
flare events ($<1$~h). As analysis method a likelihood fitting
technique has been used, performed with a model that includes the
Sun as a point source and fixed galactic and extragalactic diffuse
emission.

Moreover, the upper limit of high energy solar emission integrated
on more than one year of flares has been computed. LAT data of
flares detected by RHESSI have been collected (data selected one
hour before and five hours later with respect RHESSI flares, because
of the long duration of high energy emission). The position of the
Sun has been computed using a JPL library interface\cite{JPL} and
then data have been centered on the istantaneous solar position.
Successively these data have been merged and the analysis has been
performed, using the standard likelihood technique provided by the
LAT ScienceTools package (v9r15). Since the Sun is a moving source
in the sky, the problem is to compute the correct galactic and
extragalactic diffuse background emission as the Sun moves through
the sky. In order to evaluate the diffuse background in proper way
the fake source method has been used. The fake Sun follows the real
Sun along the same path (i.e. the ecliptic) but at an angular
distance of $30^{\circ}$. The fake Sun is therefore exposed to the
same celestial sources as the true Sun and the events observed in
the frame centered on the fake Sun make a good description of the
diffuse background. The model for the likelihood analysis is
composed by two fixed components (quiet Sun and fake Sun) obtained
in previous analysis\cite{quietSun}\cite{moriond}\cite{AIP} and the
flaring Sun free component.

\section{Results}

At 20:14:42.77 UT on 02 November 2008, \emph{Fermi}-GBM triggered
and located a very soft and bright event\cite{gcn}. The event
location was RA = 217.6 deg, Dec = -15.7 deg ($\pm 1.1$ deg), in
excellent agreement with the Sun location. The time of the event
coincides with the solar activity reported in GOES solar reports
(event 9790: onset at 20:12 UT, max at 20:15 UT, end at 20:17, B5.7
flare). This is the first GBM detection of a solar flare.
\emph{Fermi}-GBM triggered on a solar flare a second time at
19:37:46.39 UT on 28 October 2009\cite{gcn2}. LAT data have been
selected in the energy range 100~MeV - 300~GeV, according to the
solar activity detected by GOES and RHESSI: no high energy emission
has been detected by the LAT for both events.

From August 2008 to August 2009 RHESSI has detected 200 flares with
$>10^{5}$ counts. The highest energy band in which most of these
flare have been observed by RHESSI is 3-6 keV. Few flares ($<20$)
have been observed in the energy band 6-12 or 12-25 keV. Flares
outside the LAT field of view and the ones that occurred while the
LAT was transiting in the SAA have been discarded. As a result LAT
data of 80 flares have been analysed and the upper limit on the high
energy ($>100\textnormal{ MeV}$) emission has been computed for each
of these flares. No significant emission has been detected.

The preliminary upper limit on the emission of the flaring Sun
integrated over one year of flares in LAT data is $5.67\cdot
10^{-7}\text{photons cm}^{-2}\text{ s}^{-1}$. This value is derived
from a cumulative analysis of all the 80 flares with a time of six
hour around each trigger time (one hour before and five hours
later), taking into account the quiet sun component\cite{quietSun2}.
A more detailed analysis is in preparation.

\section{Conclusions}
Solar flare events have been searched in the first year of LAT data
(August 2008 - August 2009). Up until now there is no evidence of
high energy emission from solar flares detected by the LAT, while
the quiet Sun emission has been
detected\cite{quietSun}\cite{moriond}\cite{AIP}. However, the Sun is
at the minimum of its activity cycle and no intense flare has
occurred. The solar activity is expected to rise in the next months,
reaching the maximum in 2012.

We will continue€ to monitor the active regions of the Sun and to
improve our analysis techniques, waiting for an intense flare
detectable by the LAT.

\bigskip 
\begin{acknowledgments}
The \emph{Fermi} LAT Collaboration acknowledges support from a
number of agencies and institutes for both development and the
operation of the LAT as well as scientific data analysis. These
include NASA and DOE in the United States, CEA/Irfu and IN2P3/CNRS
in France, ASI and INFN in Italy, MEXT, KEK, and JAXA in Japan, and
the K. A. Wallenberg Foundation, the Swedish Research Council and
the National Space Board in Sweden. Additional support from INAF in
Italy for science analysis during the operations phase is also
gratefully acknowledged.

Work supported by Department of Energy contract DE-AC03-76SF00515.
\end{acknowledgments}

\bigskip 

\end{document}